\documentclass{aa}  
\usepackage{graphicx}
\usepackage{txfonts}
\bibpunct{(}{)}{;}{a}{}{,} 

\begin{document}

   \title{Superoscillations in solar MHD waves and their possible role in heating coronal loops.}

   \author{{\sc A.~L{\'o}pez Ariste}\inst{1,2},  {\sc M.~Facchin}\inst{1,2}}
\institute{IRAP - CNRS UMR 5277. 14, Av. E. Belin. 31400 Toulouse. France 
    \and Universit\'e de Toulouse, UPS-OMP, Institut de Recherche en Astrophysique et Plan\'etologie, Toulouse, France}
 
   \date{Received ...; accepted ...}

 
  \abstract
  {}
   {To study the presence of superoscillations in coronal magnetoacoustic waves and its possible role in heating coronal loops through the strong and localized gradients they generate on the wave.}
   {An analytic model is built for the transition between a sausage and a kink wave modes propagating along field lines in the corona. We compute in this model the local frequencies, the wave gradients and the associated heating rates due to compressive viscosity.}
   {We find superoscillations associated with the transition between wave modes accompanying the wave dislocation that shifts through the wave domain. Frequencies 10 times higher than the normal frequency are found. This means that a typical 3-minute coronal wave will locally oscillate in 10 to 20 seconds. Such high frequencies bring up strong gradients that efficiently dissipate the wave through compressive viscosity. We compute the associated heating rates. Locally, they are very strong, largely compensating typical radiative losses. }
   { We find a new heating mechanism associated to magnetoacoustic waves in the corona. Heating due to superoscillations only happens along particular field lines with small cross sections, comparable in size to coronal loops, inside the much larger magnetic flux tubes and wave propagation domain.}
  

   \maketitle

\section{Introduction}

Coronal heating continues to be, more than half a century later, one of the main problems to be solved in solar physics. Over the years, it has become clear that the source of energy to heat the corona lies in the plasma motions of the photosphere. It is also clear that this energy is transferred into the corona along magnetic field lines and that this magnetic field is also a key aspect of its deposition into the rarefied coronal plasma \citep{Parnell2012}.
    Any estimate of the plasma kinetic energy in the photosphere largely exceeds the amount of energy necessary to heat the full corona. It appears that the coronal heating problem is just a question of transferring and dissipating in the corona a fraction of this available energy. Furthermore, the low density and high temperature of this coronal plasma conspire to make it very difficult to cool  down once it reaches the 1 million K mark. {Actually radiative losses diminish with increasing temperature until reaching the 3 million K} \citep{Raymond1976,Cook1989,Lykins2013}
   Hence any  heating  that succeeds in initially overcoming radiative losses  at around several hundred thousand K will drive the plasma into higher and higher temperatures with smaller and smaller losses. This runoff will only end when the plasma reaches  3 million K, or the heating stops.     Such arguments may erroneously let think that heating the corona must  be not so hard. But, as the proverb says, the devil is in the details, and understanding the fine details of coronal heating has been proved over the years to be a difficult enquiry.

Two basic schemes are commonly accepted to solve the transfer of photospheric energy into the corona and its eventual deposition. They often go under the labels of AC (standing for Alternating Current) and DC (standing for Direct Current) mechanisms. AC mechanisms rely on MHD waves excited in the photosphere and guided into the corona by magnetic flux tubes, where they dissipate through a choice of mechanisms\citep{Arregui2015}. Magnetoacoustic waves are easily excited in the photosphere but they have been traditionally thought to be blocked at chromospheric heights. For this reason Alfv\'en waves have often been preferred as the waves that heat the corona when dissipating, often in the presence of localised phase gradients\citep{Heyvaerts1983}. Recent observations, however, have detected the presence of magnetoacoustic waves in coronal loops. While it is a subject of discussion how these waves crossed the chromospheric boundary or whether they were re-excited in the low corona rather than propagate through\citep{Khomenko2015}, it is evident that their presence in the corona allows them to participate in the task of dissipating energy and heat. Such task is hindered by their low frequency, apparently insufficient to ensure sustained dissipation rates \citep{Porter1994a,Porter1994}. 
Although the observation of Alfv\'en waves in the corona has been claimed by \cite{Tomczyk2007}, \cite{VanDoorsselaere2008}
argued that the torsional character of Alfv\'en waves would made them unobservable and that what was observed by  \cite{Tomczyk2007} was rather kink magnetouacoustic waves.  The reality may be mid way between those two claims. The presence of dislocations in CoMP data could only be explained by \cite{LopezAriste2015} through the simultaneous presence of a magnetoacoustic sausage mode {(the definition of which we give below)} with a {second }wave. This {second }wave could either be a magnetoacoustic kink   or an Alfv\'en wave. Whatever the case, the existence of Alfv\'en waves is   beyond doubt and hence they continue to play an important role in studied AC mechanisms.

DC mechanisms rely on the reconnection  of magnetic field lines. Magnetic fields are, in the photosphere, subject to plasma movements. Perhaps excluding sunspot umbrae, photospheric movements are sufficiently chaotic to braid field lines up into the corona. Reconnection of these tangled fields may  liberate sufficient amounts of energy, as can be seen during flares. Assuming that this reconnection happens at all scales, spatial and temporal,  heating the corona appears as a natural consequence. 
It is evident that both AC and DC scenarios happen simultaneously in the Sun and contribute in larger or smaller share to the coronal heating. In this sense, the problem of heating the corona has been translated into identifying those cases in which one or another mechanisms are at work, and figuring out the details of the wave transfer and dissipation, or of the magnetic field braiding or  of magnetic reconnection. In spite of that shared role, DC mechanisms  are not the subject of this work and, in the sake of brevity, we will leave this short paragraph as the only reference to them, redirecting the reader to the many excellent reviews on the subject (e.g. \cite{Rappazzo2007,Parker1988}).

The present work was not initially concerned with heating. But in the course of the investigation of the phenomenon of superoscillations in solar MHD waves it appeared that under certain conditions the high frequencies associated with superoscillations may result in high spatial gradients and hence  in an increased wave dissipation due to compressive viscosity. Such conditions appear to happen often enough in idealised coronal loops. The previous short review of the coronal heating problem frames our contribution: superoscillations may contribute to the heating of coronal loops as other known mechanisms do. Whether they heat more or less than such other AC dissipation mechanisms or DC scenarios is beyond the scope of this work. Whatever the case, even if superoscillations are able of an  important energy contribution, it would only affect a small
region in coronal loops, not  the  whole volume  of the corona.

Superoscillations, the main focus of this work, are a surprising phenomenon in many senses. They appear to be quite common, in spite of the fact that they have been overlooked in almost all studies of waves in physics until recent years. \cite{Berry2009} has mathematically demonstrated that up to 30\% of the volume occupied by a random wave field superoscillates. A third of the volume occupied by wave fields behaves therefore in a manner which has escaped the attention of scientists for years. This comes as a surprise and justifies the recent interest on that phenomenon. Can we observe those theoretical predictions in physical waves?
The next section reviews the concept of superoscillation and its discovery in observed solar magnetoacoustic waves.

As its name hints, superoscillating waves present local frequencies much higher than expected (the actual quantitative definition is given in Section 2.1). Having been seen in solar magnetoacoustic waves, it is natural to wonder whether such higher frequencies dissipate the wave and if so, efficiently enough to overcome radiative losses. We address here that question analytically and we anticipate the answer of Section 3: under appropriate conditions they do heat enough. Those appropriate conditions are perhaps too idealistic and Section 4 will be dedicated to list and discuss the most important approximations made and whether it can be justified to use them in coronal waves.

\section{Superoscillations}

\subsection{Definition and mathematical aspects}

Superoscillations have been defined (discovered) by \cite{Berry1994} following a suggestion by \cite{Aharonov1990} as those regions in a band-limited wave field where the local frequency is higher than the highest frequency of its Fourier spectrum. Given such a definition, it is not surprising that superoscillations have escaped the attention of wave studies so long. Waves are almost always studied through its spectrum, through its Fourier components, which, individually, carry no information on particular places where the wave happens to oscillate faster than any of its components.

In trying to explain how it is possible that a wave oscillates faster than any of its Fourier components, one should start by defining the  local frequency. Given a wave propagating in the $z$ direction and described by a function $f(z-ct)$, where $t$ is time and $c$ is the phase speed of the wave, we define a local frequency as the temporal derivative of the phase of the wave. That is, we write the wave function in polar form,
$$f(z-ct)=\rho (z-ct) e^{i\chi(z-ct)}$$
with $\rho$ the scalar amplitude of the wave, and $\chi$ its phase. The local frequency is given by
$$\hat \omega = \frac{\partial}{\partial t} \chi(z-ct)=\partial _t \chi(z-ct)$$
where we introduce a self-explanatory short-hand convention for the derivative.  In the particular case of a plane wave (or any isolated Fourier component)
$$f(z-ct)=Ae^{ik(z-ct)}$$
and the local frequency is constant and equal to $\omega=kc$ the global frequency.
A similar definition can be used with  the wavenumber, if deriving with respect to the spatial coordinate $z$; a local wavenumber appears and superoscillations can also be associated with this local wavenumber. 
Both can be readily generalized for 3-dimensional waves propagating in one or several directions.

As said above, in a single monochromatic plane wave, no local wave frequency can be higher than the global frequency . One needs what \cite{Berry2012} have described as a \textit{delicate conspiracy of phases} during the  destructive interference of several plane waves to generate a superoscillation. This description brings forward two different aspects of superoscillations: they require complicated waves and they appear near places where the amplitude of the wave is greatly diminished due to destructive interferences. \cite{Kempf2000} has given a mathematical demonstration of the relation between superoscillations and exponentially decreasing amplitudes in terms of Shannon theorem and information theory. In strawman parlance:  one can have a wave superoscillating beyond its global frequency, but at the price of a vanishing amplitude. One can even find an asymptotically infinite frequency, but  in such place the wave amplitude will be  rigorously zero.

\begin{figure*}
   \centering
   \includegraphics[width=\textwidth]{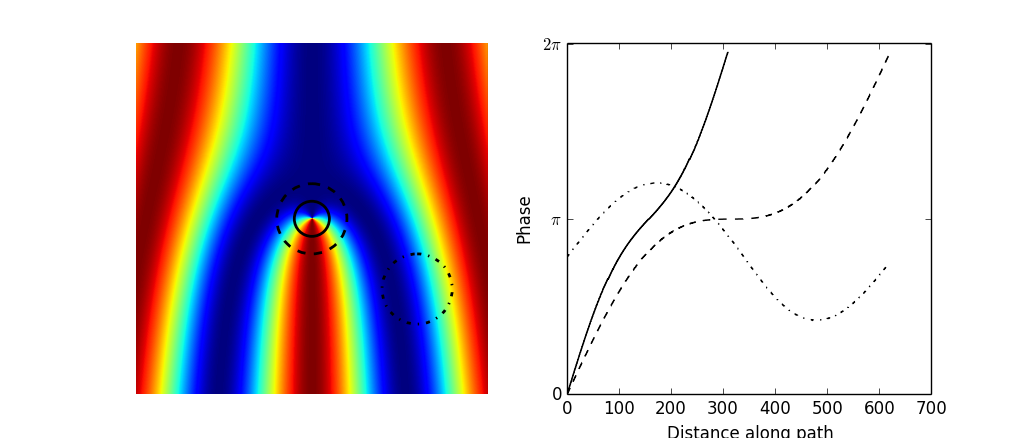}

      \caption{Left: A scalar wave carrying a dislocation is imaged as the real part of the function $(y+ix)e^{ix}$ in the plane $(x,y).$ Only a zoom over the region around the dislocation is shown. Three monodromies are drawn, two as circumferences centred at the dislocation and one that does not enclose it. The phase change along these paths is $2\pi$ if the singularity is enclosed, but zero if it is not. Right: the phase along the three monodromies. Because of its shorter length, the phase undergoes a more rapid change for the inner monodromy enclosing the dislocation.}
         \label{scheme}
   \end{figure*}

This correlation between superoscillations and low amplitudes brings up another aspect of this phenomena: it usually appears in the neighbourhood of wave dislocations. Wave dislocations are phase singularities: the phase is undefined and the amplitude is zero. As such, along a closed path around a singularity (a curve called a monodromy), the gradient of the phase does not add up to zero but rather to an integer multiple of $2\pi$.  That is, the phase changes by a multiple of $2\pi$ when travelling along this path (see Fig.\ref{scheme}). The monodromy can be as near to the singularity as desired and it may have a very short length.  Thus the gradient of phase as a function of length along the monodromy can be arbitrarily high. But such monodromy path is just a mathematical construct to put in evidence the existence of the singularity. It has no \emph{ a priori} relationship with the direction of propagation of the wave. But let us suppose for a moment that it does, and that the length along the monodromy can somehow be made to correspond to $z$ or time. It  is then clear that the proximity of  a dislocation brings up superoscillations. The previous description may give the impression that the appearance of superoscillations requires 
tough conditions: presence of a dislocation, short paths around it that coincide with $z$ or time. But this is not so. \cite{Berry2009} have demonstrated that in a volume where plane waves travel in random directions and interfere, up to one third of it is superoscillating, particularly in the neighbourhood of the large number of dislocations present. 

One third of the volume of a wave field superoscillating is a disturbingly large volume for a phenomenon that has escaped the attention of wave studies until recently. A few non-rational reasons for this oblivion: first, it is not uncommon to start by decomposing any wave in its Fourier components which are subsequently referred to as modes and studied independently; second, all interactions between modes are often assumed to be the result of non-linear processes; third, the apparent presence of dislocations and local frequencies that depart from the previous scheme are disregarded as \textit{accidents} due to particular boundary conditions. 

As mentioned above, the Fourier modes cannot capture the superoscillation which, by definition, locally oscillates faster than any Fourier frequency. Hence any Fourier decomposition will miss the superoscillations. 
It is also important to stress that superoscillations (and for the sake of it, dislocations) are not due to any non-linearity in the wave or in its evolution. They are perfectly stable solutions of the linear scalar wave equation. It is however true that the higher the superoscillation the lower the wave amplitude, so it is legitimate to think of superoscillations as higher-order perturbations to the wave dynamics and evolution.


 Beyond solar physics, the relatively recent discovery of light vortices \citep{Allen2002} carrying wave dislocations has brought this field into the light of research in optics and its applications. Superoscillations of electromagnetic radiation have been used in the radio domain to create highly collimated beams \citep{Berry2014}. These few examples  of superoscillations and dislocations elsewhere in physics carries the message that one and the other are not just mere minor accidents of the wave field due to particular boundary effects, but that they can have an impact in the evolution of the wave field. Our purpose here is to analyse whether this conclusion can be also applied to magnetohydrodynamic waves in the solar atmosphere.

\subsection{Superoscillations in solar magneto-acoustic waves: observation and modelling}

In most of the mathematical literature cited above, superoscillations are built from a small set of  well-known analytic functions that have the desired mathematical properties. Though such analytical functions are  mathematically rigorous solutions of a wave equation, they are not, in general, solutions naturally excited in the conditions that one can expect in the solar atmosphere. A better, though perhaps tortuous, path towards superoscillations in solar magnetoacoustic waves is the one described in the previous section: a path around a dislocation superoscillates.

Hence we shall start this section by reviewing  the discovery of  dislocations in solar magnetoacoustic waves, both observationally and theoretically \citep{LopezAriste2013,LopezAriste2015,LopezAriste2016}.  A simple model of waves propagating along a vertical magnetic flux tube in the photosphere of the Sun is sufficient to make dislocations appear in the analytical solutions. Such vertical flux tube constitues a cylindrical geometry on which waves are 
excited to travel preferentially along the field. In such conditions  the transverse structure of the natural solutions to the wave equations is written in terms of the Bessel functions times an azimuthal dependence. The basic mode, with no azimuth dependence, is written in terms of the $J_0(r)$ Bessel function and is often referred to as the sausage mode. The next mode, has an $e^{i\theta}$ azimuth dependence and a radial dependence in terms of $J_1(r)$ and is often called, the kink mode. 
{ In what follows we shall use these two terms, sausage and kink, to refer to the velocity vector solution to the wave equation with those basic azimuth dependences. The waves we are going to call sausage or kink, both have longitudinal and transverse velocity components, although we will focus in what follows mostly in the longitudinal component of both sausage and kink velocity waves, and we will distinguish them by their azimuth dependence. Looking into this  azimuth, $\theta$, we realise that it} is not defined at $r=0$ and therefore the kink mode carries a singularity, a dislocation, at this point. This has been observed directly in the velocity component  along the magnetic field (along $z$) of magnetoacoustic waves observed in the umbra of sunspots at both photospheric and chromospheric heights \citep{LopezAriste2016}.   

 In the time series that makes up those  observations, both a sausage mode and a kink mode were observed sequentially. The presence of the dislocation is what allowed the authors to identify unmistakably one and the other modes. It was a surprise to  see that, in those observations of waves in the solar photosphere, one mode can be excited independently of the other. This means that the excitation mechanism is not a bare introduction of energy.  A diagram of possible kink and sausage modes in photospheric conditions as a function of phase speeds and wave number (see for example Fig. 3 of \cite{Edwin1983})shows that, for slow modes in the photosphere, in the vicinity of every allowed sausage mode there are many other sausage and kink harmonics with almost identical speeds and wavenumbers. Based upon those diagrams one expects that upon excitation, many different kink and sausage modes are simultaneously excited, not just one of them as the observations show.
 
 Even more surprising is the fact that the excitation mechanism was also capable of discriminating the two possible azimuth dependences of the kink mode. In a kink mode,  both the $e^{+i\theta}$ and the $e^{-i\theta}$ dependencies are valid and energy-degenerate solutions of the wave equations. Being energy degenerate, it is natural to assume that both are simultaneously excited with equal amplitudes. The result would be plasma that oscillates in the left-right direction in a vertical field, a movement that gives the kink mode its name (left and right are here arbitrary  directions in the plane transverse to the propagation).  But this is not what was observed by \cite{LopezAriste2016}. Those authors demonstrated that only one of the polarizations had been excited (the $e^{+i\theta}$ in their sign convention) resulting in a helical wavefront \citep{Sych2014}.
 This chirality preference among energy-degenerate solutions is a signature of a non-trivial excitation mechanism for those magnetoacoustic waves in sunspots.

But the one result in \cite{LopezAriste2016} of interest for the present work  stemmed from the detailed study of how the kink mode did substitute the sausage one in the span of about one period. Neighbouring pixels which during the sausage mode oscillated in phase ended up oscillating in phase opposition as the singularity of the kink mode settled between them. The transition from in-phase oscillation to phase opposition was seen, both observationally and analytically, to happen through an increase in the local frequency of the wave: a superoscillation.   At the beginning of this section we mentioned that our path to find superoscillations would require traveling along the path around a dislocation. Rather than having plasma moving along arbitrary mathematical paths, we now see that during the transition from one wave mode neighbouring plasma has to oscillate 
in drastically different and quickly changing phases and frequencies as the dislocation moves in its close proximity.

This is therefore the basic scenario that we will in what follows model analytically to study superoscillations in solar waves:  a sausage mode propagating vertically is substituted by a kink mode in the span of one period or less. To that end,
we will model the atmosphere as a vertical magnetic flux tube of diameter comparable to (but smaller than)  the size of the umbra of a sunspot (several tens of arcsec). In advance of next section, we underline the fact  that the size of this magnetic flux tube is much larger than the size of coronal loops.
 Inside this flux tube a magnetoacoustic wave is excited. { The transverse extension of this excitation is supposed to be the whole flux tube.} Observation of waves at chromospheric heights show that the initial excitation region in this transverse, horizontal plane is much smaller than the flux tube  \citep{RouppevanderVoort2003,Bogdan2006,Sych2014} and that the propagation along the magnetic field lines is accompanied with a transverse propagation. In spite of that we will simplify the problem and assume that the wave is excited in the full magnetic flux tube and propagates exclusively along the field lines. We shall call  $z$ this direction; $r$ and $\theta$ will complete the cylindrical coordinates used in the rest of this paper. { We will further assume that this flux tube continues vertically into the corona, and that the excited waves propagate through this whole volume, much larger than single coronal loops.}

Among the several works in the literature that solve the wave equation in such conditions, we shall use the conventions of \cite{Edwin1983} and recover for the magnetoacoustic wave two modes with different  azimuthal dependence $e^{im\theta}$. {\text As said above we shall use the terms sausage and kink to refer to them:} a sausage mode with $m=0$ and no azimuthal dependence, and a kink mode with $m=\pm 1$. 
 In the literature one can find as well the use of {\em body waves} and {\em surface waves} to refer to those solutions \citep{Edwin1983}.  {We stress once again that our so defined sausage and kink modes have both longitudinal and transverse components. Indeed }
 explicit expressions for all  the vector components of the velocity field $v_r$, $v_{\theta}$ and 
$v_z$ { of both these kink and sausage modes} have been given by \cite{LopezAriste2015} in the case of our simplified vertical magnetic field with exclusive propagation along the field. But observationnally it is in $v_z$, { of both the sausage and kink modes} that superoscillations have been observed so we will restrict our analysis to  this component. Undoubtedly, superoscillations will exist also in the other velocity components but, first, we wish to reproduce what has already been observed, and, second, changes in frequency of $v_z$ appear to have a larger impact on wave dissipation than similar changes in the transverse components  \cite{Porter1994a}.

The vertical velocity component $v_z$ of the wave is explicitly given by
\begin{equation}
v_z(r,\theta,z,t)=-iA\frac{c_s^2k}{\omega^2}J_m(-m_0r)e^{im\theta}e^{ik(z-ct)}
\label{Eq1}
\end{equation}
Letting $m=0$ we recover the sausage mode, and with $m=1$ the kink mode. $A$ is a scalar amplitude common to the three velocity components (though potentially different for the kink and the sausage modes); $c_s$ is the speed of sound in the atmosphere, $\omega$, $k$ and $c$ are respectively the general wave frequency, wave number and phase speed of the wave. We fix $\omega$ to correspond to a 3-minute period wave which can potentially  cross the chromospheric cut-off and propagate into the corona. The dispersion relation provides several solutions for $k$ and $c$, even after fixing the frequency, depending in particular on $m$. But for all of them, the phase velocity is bounded between the speed of  sound $c_s$ and a characteristic velocity $c_T$ given by 
$$c_T=  \frac{c_sv_A}{\sqrt{c_s^2+v_A^2}},$$
where $v_A$ is the Alfv\'en speed.  For our numerical examples below we shall pick the average of those two velocities, $c_s$ and $c_T$, as a representative phase velocity for both wave modes. Such choice also fixes the value of  the wave number $k$.

The transverse scale of the wave is fixed by the parameter 
$$m_0^2=\frac{(k^2c_s^2-\omega^2)(k^2v_A^2-\omega^2)}{(c_s^2+v_A^2)(k^2c_T^2-\omega^2)}.$$
 In typical photospheric conditions, setting $c_s=10 km/s$ and $v_A$ twice that value, $m_0$ sets a scale of $1600 km$ or about 2-3 arcsecs. Given those values and with our particular choice of the phase velocity, the first zero of the $J_0$ Bessel function will be found at 5-7 arcsecs of the center of the wave. In the corona, where the speed of sound is 10 times larger, and the Alfv\'en speed approaches $1000 km/s$  the transverse scale of the wave roughly doubles.  Evidently it would be sufficient to pick another value for the phase velocity to make those transverse scales bigger and comparable, for example, with the size of the umbra in the photosphere. But since this scale is not of particular importance in the present work, we will stick to the adopted value of the phase speed and will limit the lateral size of the oscillating region to 1 or 2 times $1/m_0$.

As said in the first paragraphs of this section, the kink mode has a phase singularity at $r=0$ where $\theta$ is not defined. The amplitude of the wave at that point is  $J_1(0)=0$. The sausage wave on the other hand has its larger amplitude at precisely that point. Hence, the transition from sausage into kink results in a substitution of the maximum amplitude of the wave by a zero-amplitude singularity at $r=0$.  We will analytically model this transition with an amplitude modulating function $\beta(z-ct)$ that varies from 0 through 1 roughly linearly. The dependence of $\beta$ on $(z-ct)$ will ensure that the combined wave still is a solution of the wave equation. Explicitly, in order to make this function derivable at all times, we define
\begin{equation}
\beta (z-ct)=\frac{1}{1+e^{-s(kz-\omega t+\phi)}}
\end{equation}
The slope parameter $s$ allows us to control how long takes the  transition: the $\beta$ function, thus defined, 
changes from 0.01 to 0.99 in  $\ln(99)/(\pi s)$ wave  periods.  A phase $\phi$  will help  fixing when the transition starts.

The final expression for $v_z$ transitioning from sausage to kink is then given by
\begin{eqnarray}
v_z(z-ct)=-i\frac{c_s^2k}{\omega^2}[(1-\beta)A_k J_1(-m_0r)e^{i\theta}e^{ik_1(z-c_1t)}+\nonumber \\
 \beta A_sJ_0(-m_0r)e^{ik_0(z-c_0t)}]
 \label{vz}
\end{eqnarray}
where $A_k$ and $A_s$ are the scalar amplitudes of the kink and sausage modes, and $k_1$ and $k_0$ ($c_1$,$ c_0$) are 
the potentially different wave numbers (phase speeds) of those two modes.  In the sake of simplifying the  coming expressions, it is convenient to define
$$\rho_0=\frac{c_s^2k}{\omega^2} A_sJ_0(-m_0r)$$
(and analogously $\rho_1$) and 
$$\Delta=k_1(z-c_1t)-k_0(z-c_0t).$$
Since the local frequency is defined as the derivative of the phase of the wave, the previous Eq.\ref{vz} for $v_z$ is better  rewritten in polar form:
$$v_z=\rho e^{i\chi}e^{ik_0(z-c_0t)}$$
The local frequency is then given by
\begin{eqnarray}
\partial_t \chi=-k_0c_0+\Delta\frac{(1-\beta)^2\rho_1^2+\beta(1-\beta)\rho_1\rho_0\cos (\theta+\Delta)}{\rho^2} \nonumber \\
-\partial_t \beta \frac{\rho_0 \rho_1\sin(\theta+\Delta)}{\rho^2}
\label{super}
\end{eqnarray}

Of the three terms in the right-hand side of Eq.\ref{super}, the first one is just the global frequency $\omega$ of the sausage mode while the second ensures that, as $\beta$ changes from 0 to 1, the global frequency changes from that of the sausage to that of the kink, if there is any difference among them. 
With the first two terms the local frequency is  bounded between the global frequencies of the kink and the sausage, and it is constant if those two frequencies are equal, which shall be the case in all our numeric examples. The third term brings in the superoscillations: the larger the derivative of $\beta$, the larger this term is, and it will get even larger if we approach a place where $\rho=0$, a dislocation. Unlike the second term, this superoscillation term diverges when $\rho=0$.  This singularity is not coincident with the singularity of the kink mode ($\rho_1=0$ at $r=0$) nor with the virtual singularity $\rho_0=0$ at $r=\infty$ of the sausage mode.  We can compute the position of this dislocation  $\rho=0$ as a function of  $\beta$ by solving the two equations
\begin{eqnarray}
\theta+\Delta=0 \, \mod \, \pi \\
(1-\beta)A_1J_1(m_0r)+\beta A_0 J_0 (m_0r)=0.
\end{eqnarray} 
As time goes on and $\beta$ changes, the position of the singularity $\rho=0$ also varies and, in its proximity, the wave superoscillates.

\begin{figure*}
   \centering
   \includegraphics[width=\textwidth]{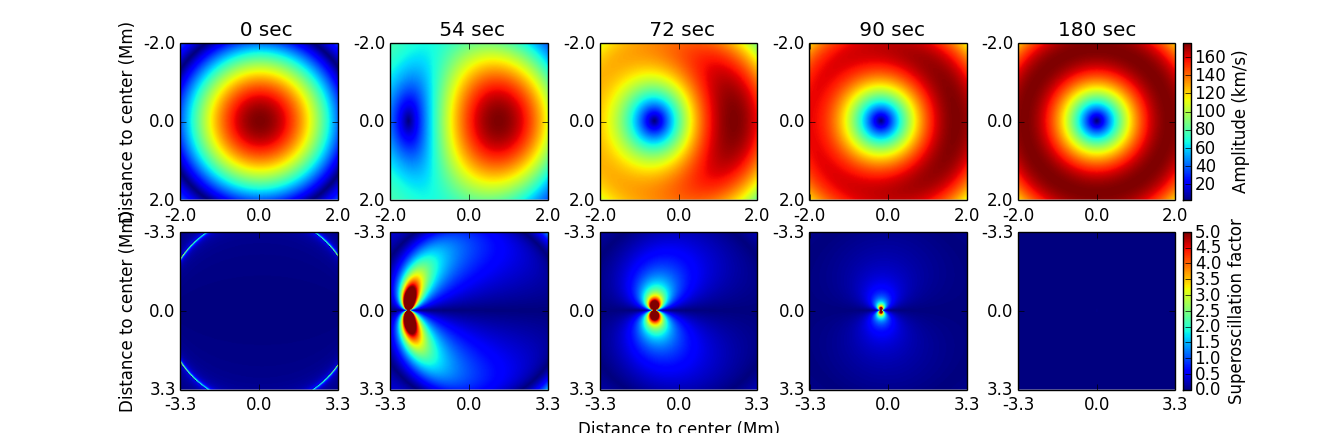}

      \caption{Amplitude (top) and superoscillation factor (bottom) of the magnetoacoustic wave in the plane transverse to propagation at a fixed height as it transfroms from sausage (left) to kink (right). Details on the parameters of the model are given in the text.}
         \label{F1}
   \end{figure*}

Figure \ref{F1} shows a particular realisation of the previous expressions for the case of the sound speed being 100km/s, the Alfv\'en speed 10 times larger, and the phase speed the average of $c_s$ and $c_T$, identical for both wave modes. The scalar amplitude of the kink is set to $0.2 s^{-1}$ and that of the sausage to $0.1s^{-1}$. With these values, the transverse amplitude of the velocity vector is 1.5km/s, comparable to the measurements of CoMP in coronal loops \citep{Tomczyk2007,Threlfall2013}. During the transition from sausage to kink, the singularity $\rho=0$ appears around time $t_0$ at the outer limit of the wave domain and approaches monotonically the final position of the kink singularity at $r=0$. In Fig.\ref{F1} we plot  at five different times during this transition both the wave amplitude (which allows one to identify the sausage and kink amplitude distributions at both ends and the mixed solutions during the transition) and  the ratio of the local frequency $\frac{\partial \chi}{\partial t}$  respect to the global frequency $\omega$ of this 3 minute-period wave. During the transition,  the local frequency is larger than $\omega$ (values in the bottom plots of Fig.\ref{F1} larger than 1) at many places, reaching easily 10 times that value around the singularity. A local frequency 10 times  higher than the global one means that, locally, the combined wave is oscillating with a period of just 18 seconds.

The simplifications introduced in our model allow us to compute analytically this superoscillation during the transition from sausage to kink. This was important in order to convey the message that superoscillation is a property of these waves and not the result of numerical issues during the modelisation of the wave. The superoscillation appears during the transition from a sausage ($m=0$) to a kink ($m=1$) magnetoacoustic wave mode, as the dislocation, and the rapid phase changes in its proximity, travel through the wave domain. The faster this transition is, the larger the superoscillation. 

That these transitions between wave modes happen is obvious. Beyond this fact is the stronger one that such transitions have been observed in the longitudinal velocity component of waves in the umbra of sunspots, at photospheric and chromospheric heights. Preparing the way for the next section, we shall assume that such transitions also take place at coronal heights. Beyond any discussion on how such waves can cross the chromospheric boundary, dislocations in coronal waves have been observed \citep{LopezAriste2015} and they have been  modelled also as the superposition of a kink and a sausage wave modes. In one way or another, the magnetoacoustic modes observed in the chromosphere, both kink and sausage, travel into the corona. We should also expect in the corona the superoscillations associated to the transition between one mode and the other.

\section{Viscosity and heating associated to superoscillations}

Viscous heating is associated with gradients. \cite{Heyvaerts1983} proposed phase mixing as a source of phase gradients resulting in Alfv\'en wave damping in the corona and, eventually, heating. \cite{Porter1994a,Porter1994} demonstrated that phase gradients associated with the mode spatial distribution of amplitudes resulted in compressive viscous disipation in magneto-acoustic waves. However, the typical frequencies of coronal waves were not sufficient to compensate radiative losses. Higher frequencies could increase the compressive viscosity and overcome radiative losses, heating the corona \cite{Porter1994a}.  Superoscillations with their high local frequencies  could provide in a natural manner the strong gradients needed for efficient wave damping. But, as we saw in the previous section, superoscillations occur in coincidence with  dislocations and their associated  drastic fall of amplitude.  The prime question we address in this section is whether such low amplitudes around dislocations are nevertheless large enough to dissipate enough energy through the strong gradients that superoscillations bring.
In order to answer such question, having an analytical model, in spite of the simplifications, is a must: we can clearly discern heating due to superoscillations from other sources of heating.

As discussed by \cite{Porter1994a} the coronal plasma falls in the strong-field regime of \cite{Braginskii1965}.  In this regime the fluid is non-newtonian and one needs to explicitly compute the stress tensor in the momentum equation, rather than relying on a by-default wave damping through  the laplacian of the velocity.  \cite{Braginskii1965} actually gives an explicit expression for the stress tensor $\Pi_{\alpha \beta}$ in the strong-field regime. But rather than solving the momentum equation and look for an eventual wave damping, we are going to use such an  explicit expression directly in the expression for the heating rate. They are given by 
$$Q_{viscous}=-\Pi_{\alpha \beta}\frac{\partial v_{\alpha}}{\partial x_{\beta}}$$
where the $\alpha$ and $\beta$ indices run over the three spatial coordinates, and the Einstein summation condition applies.
Four different viscosities appear in the stress-tensor and in the viscous heating rate $Q_{viscous}$  above; but one of them, the compressive viscosity $\mu_0$, is several orders of magnitude larger than the others in coronal conditions. Reduced to the term involving this compressive viscosity, the heating is explicitly written as
\begin{equation}
Q_{\mu_0}=\frac{\mu_0}{3}\left(\partial_r v_r+\frac{1}{r}\partial_{\theta} v_{\theta} +\frac{1}{r}v_r-2\partial_z v_z\right)^2
\end{equation}
While there is no further difficulty in numerically computing this heating from our analytical solution for the velocity vector, it is worth  further simplifying  the above expression  to gain some insight on the different heating sources . Among them, we expect to find superoscillations and we aim at comparing the heating they produce with what other heating sources may do. As we saw in Eq.\ref{Eq1}, the amplitude of the $v_z$ component depends on two scalar factors: one, dubbed $A$ is a scalar amplitude common to the three components of the velocity vector; the second 
$$\frac{c_s^2k}{\omega^2}$$
affects only  the $v_z$ component, while both the $v_r$ and $v_{\theta}$ components are multiplied by the factor (see the explicit solution in Eqs. (7) and (8)  of \cite{LopezAriste2015})
$$\frac{\omega^2-k^2c_s^2}{\omega^2m_0^2}.$$
For the typical values of frequency $\omega$ and speeds of sound ,$c_s$ and Alfv\'en $v_A$ in the solar corona, this second factor is one order of magnitude smaller than the one multiplying $v_z$.  Furthermore, the gradients respect to $r$ and $\theta$ are seen to be small compared to those respect to $z$ . All these arguments suggest that most of the expected heating, superoscillatory or not, will come from the $v_z$ component alone { or rather from its gradients along the field.}. This is helpful to our purposes in the sense that it is in $v_z$ that superoscillations have been observed.

To continue our analysis of the heating arising from superoscillations, we can approximate the compressive viscous heating to 
\begin{equation}
Q_{\mu_0}\approx \frac{\mu_0}{3}(\partial_z v_z)^2.
\label{Q}
\end{equation}
Given $v_z$ in polar form
$$\partial_z v_z = \left(\partial_z\rho +i\partial_z \chi \rho +i k_0\rho \right) e^{i\chi}e^{ik_0(z-c_0t)}$$
We introduce at this point our particular wave transitioning from sausage to kink through the $\beta(z-ct)$ function (Eq. \ref{vz}); and we collect the terms of $\partial_z v_z $ above in two groups: A function $g$  multiplying $\partial_z \beta$ and a function
 $f$ independent of it:
$$\partial_z v_z = f+g\partial_z \beta .$$
The results of \cite{Porter1994a,Porter1994} insist in that gradients of magnetoacoustic waves at typical coronal frequencies, given by the function $f$ above, are not sufficient to overcome radiative losses. If any net coronal heating is to be found, it will be attached to the transition between sausage and kink modes described by the $g\partial_z \beta$ term. 
The $g$ function has the explicit expression
\begin{eqnarray}
g=-\frac{\rho_0\rho_1\sin(\theta+\Delta)}{\rho^2}-\frac{\rho_1^2-\rho_1\rho_0\cos(\theta+\Delta)}{\rho}+
\nonumber \\
\beta\frac{\rho_1^2+\rho_0^2-2\rho_1\rho_0\cos(\theta+\Delta)}{\rho}
\label{g}
\end{eqnarray}
We recognise in the first of those three terms in Eq.(\ref{g}) the expression for the superoscillation found  in Eq.\ref{super}. The other two terms provide the gradient due to the changes in the amplitude of the wave, not of its phase. These two last terms diverge around the singularity $\rho=0$ linearly, while the superoscillatory term diverges quadratically. We expect therefore that around the singularity  the function $g$ will be dominated by the phase gradient, that is, by the superoscillation.    Introducing 
the expression of $\partial_z v_z$ in Eq.(\ref{Q}) we find
\begin{equation}
Q_{\mu_0}\approx \frac{\mu_0}{3}\left(f^2+\partial_z \beta ^2g^2+2fg\partial_z \beta\right).
\end{equation}
where the superoscillations-dominated $g$ is weighted by the steepness of the transition $ \partial_z \beta$ and added to the heating source studied by \cite{Porter1994a,Porter1994} and others.

Summarising the previous results, a wave transitioning from sausage to kink has an extra gradient term in the expression of 
$\partial_z v_z$ dominated by the superoscillations in the neighbourhood of the dislocation. This extra term was not present in 
previous studies of coronal heating associated with the compressive viscosity in magnetoacoustic waves. It is time now to compare the relative magnitude of the new and old gradient terms and determine whether the new heating source contributes any sensible heating.

\begin{figure*}
   \centering
   \includegraphics[width=\textwidth]{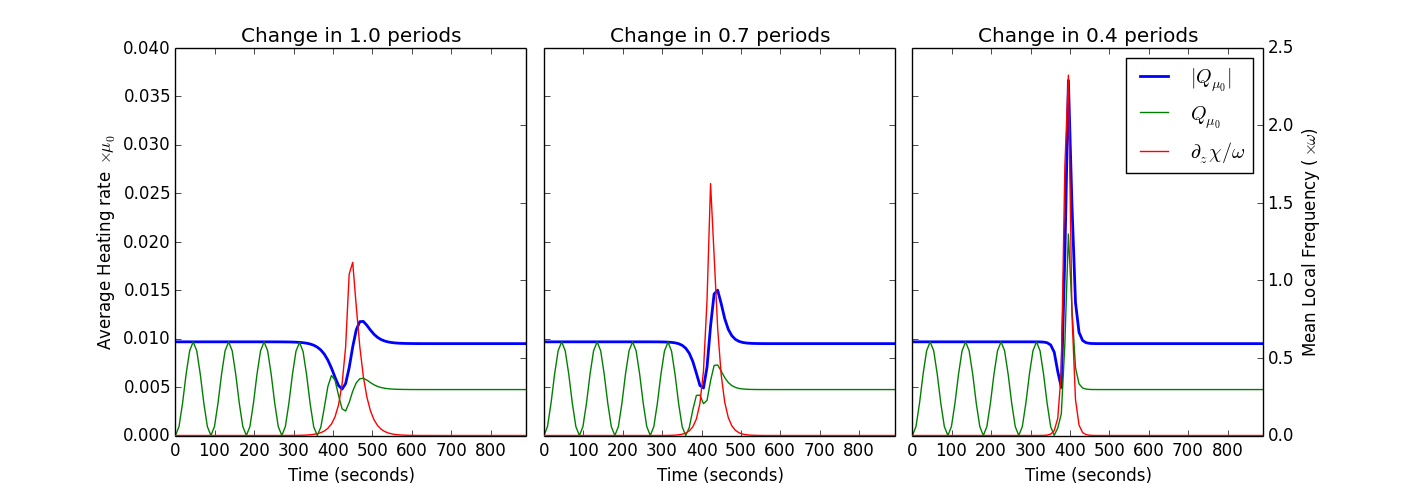}

      \caption{Plots of heating integrated in a transverse plane as a function of time for three different cases in which the change
      from sausage to kink takes place in, respectively, 1, 0.7 and 0.4 periods. The green curve is the actual heating, which depends on the phase of the wave. The blue curve is the envelope of this heating for all possible phases. It represents the maximum heating attainable. The red curve (right axis) shows the mean superoscillation factor. The highest heating rates coincide in time with the highest superoscillation. }
         \label{Total_Q}
   \end{figure*}

We have given before some of the characteristic values in the solar corona for the model parameters: the frequency is set to the 3-minute characteristic period, the speed of sound to 100 km/s and the Alfv\'en speed to 1000km/s.
The dispersion relation provides several solutions for the phase velocities of the different modes. All of them lay between the speed of sound, $c_S$, and the characteristic speed $c_T$. We chose for our figures the average of those two velocities as the phase speed of both wave modes in all our computations. We still have to determine  the steepness $s$ of the $\beta$ function that fixes how fast the kink wave substitutes the sausage wave.
\cite{LopezAriste2016} observed that the transition in the chromosphere happens in less than one period. With that constraint we are going to examine three cases: substitution in 1, 0.7 and 0.4 periods. In Fig.\ref{Total_Q} we plot the value of $Q_{\mu}$ integrated at a fixed height $z$  as a function of time (up to a factor $c$ in the scale, this figure  could also be interpreted as a fixed time at different heights $z$). After roughly 2 periods (360 seconds) the kink mode starts substituting the sausage. 

In the sausage mode, which dominates the wave up to 350 seconds, the whole transverse plane has the same phase. The amplitude is periodically 0 or maximum for the full plane altogether. The associated heating is either 0 or maximum respectively with the same period. The situation is completely different for the kink mode, which starts after 450 seconds. The kink mode, with a well defined and unique sign for the azimuthal dependence $m$, is a vortex wave. At any time in a transverse plane, two regions at opposite directions will show maximum amplitude, and two regions, at 90 degrees from the previous ones, will show 0 amplitude.  As the wave advances, such regions will rotate, but they will always be present. In consequence the heating integrated over the plane will be a constant. This is what we see in the plots after 450 seconds.

If the transition between the two modes starts while the sausage mode is near a maximum of amplitude, there will be a lot of sausage wave amplitude to be dissipated. Similarly if the transition happens when the maximum amplitude of the kink mode coincides with the region of maximum potential heating, there will be a lot of kink amplitude to be dissipated.  In summary, the actual heating will depend on the actual phases of the sausage and kink modes during the transition. In Fig. \ref{Total_Q}
we have made a suitable choice of these phases that results in a maximum heating for  the case of a transition in 0.4 periods, the plot on the far right of the figure. But such choice is not optimal for the other two cases. This particular example is represented by the green thin curve. With other choices of phases, the heating changes and it can be smaller or larger. We have computed, at every time step, what is the maximum heating that could be obtained were the phases appropriately chosen. This envelope of maximum heating is shown in the blue thick line of Fig. \ref{Total_Q}.  The third line in these plots, red in colour, shows for reference the average local frequency over the plane, in units of the global frequency $\omega$.

The most obvious feature of these three plots is the appearance of a heating pulse which coincides in time with the superoscillation. This pulse starts being noticeable when the transition happens in one period and grows in size for shorter and shorter transitions, proportionally to the larger and larger superoscillations. This is particularly clear in the blue thick line representing the maximum heating possible. For particular choices of the phases (of which the green thin line is one example) the pulse may or may not be there. In our example of a transition in 1 period, we see that the maximum heating is slightly larger than the average heating, but the phases of the particular wave plotted in green are such that it misses this maximum possible heating. Exactly the opposite happens in the example of a transition in 0.4 periods. The phases of the particular wave plotted are such that it fully experiences the heating associated with the superoscillation.   

Whenever the transition happens in less than one period, the possibility of a heating pulse associated with the superoscillation appears. The oscillating plasma has to be in phase with that pulse to actually show heating. The magnitude of this heating is several times larger than the background heating associated with either the kink or the sausage modes independently. 
\begin{figure*}
   \centering
   \includegraphics[width=\textwidth]{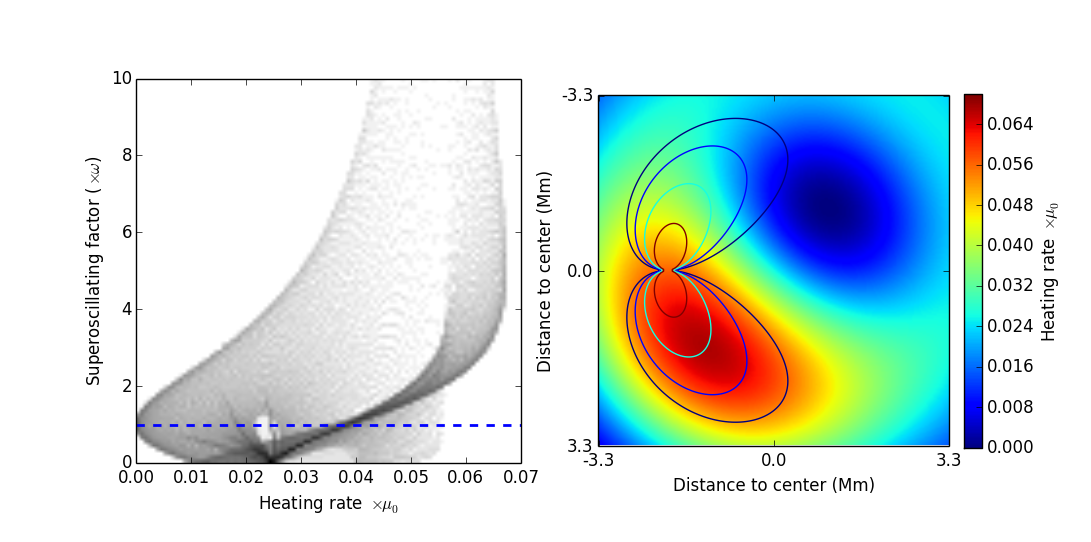}

      \caption{Spatial distribution of heating rates and superoscillation in the plane transverse to the propagation during the transition from sausage to kink. At left a density plot of superoscillations vs. heating rates where this later is seen to reach a maximum for superoscillation factors of around 3. At right a map of the transverse plane with heating rates in color and superoscillation factors in contours (values of 2,3,5 and 10 times the global frequency shown). Two superoscillatory lobes appear around the position of the dislocation, but only one of them is associated with high heating rates. This asymmetry is due to the interaction between the $f$ and $g$ parts of the expression for the heating, see text.}
         \label{HistoT}
   \end{figure*}

\begin{figure*}
   \centering
   \includegraphics[width=\textwidth]{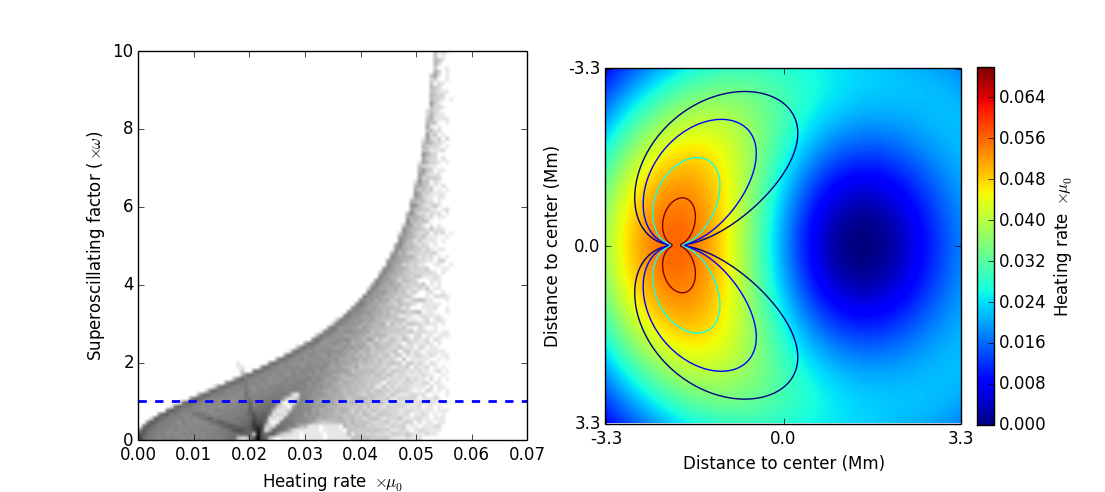}

      \caption{As in Fig.\ref{HistoT}, the spatial distribution of the heating rates is shown for exactly the same conditions, but this time only the $g$ part of the expression for the heating is shown. This $g$ function only takes into account the superoscillatory contribution and, as expected, it is perfectly cospatial with the superoscillation contours.}
         \label{HistoG}
   \end{figure*}
  
 After examining the coincidence in time of the superoscillation and the heating pulse, we look at the spatial correlation. This is 
 what    Figures \ref{HistoT} and \ref{HistoG} reveal. Both figures plot the spatial distribution of the heating at the time of the pulse associated with the superoscillation for the case of a transition in 0.4 periods. In both figures, the left plot shows a 2-dimensional histogram in terms of heating rate per volume element and local frequency. For the purpose of this histogram, the transverse plane of the wave at fixed $z$ and $t$ has been pixelized in a $200\times 200$ matrix. The right image of both figures shows the heating rate per volume element and, superposed, a contour plot of the superoscillating factors (levels at 2,3,5 and 10 times $\omega$). In Fig \ref{HistoT} the total heating $Q_{\mu_0}$ is shown. The histogram in this figure shows that, indeed, there is a correlation between higher local frequencies and higher heating rates. However, this correlation is not monotonous. The image on the right unveils the reason: the highest heating rates are not exactly superposed to the highest local frequencies. There is an apparent displacement between one and the other. Of the two lobes of high local frequencies around the singularity, only one appears to heat, while the other lies outside the heating zone. The reason for this, has 
 actually been discussed before: Superoscillations heat through the $g$ function , but to the total heating also contributes the 
 $f$ function associated with the amplitude gradients and, most importantly, the cross factor $2gf$ which is actually responsible of this displacement seen in Fig.\ref{HistoT}. Fig. \ref{HistoG} plots the heating due to $g$ alone, and here the concordance between heating and superoscillation is complete.  Once again the relative phases between sausage, kink and the $\beta$ function are critical. Heating will increase when the phases make the locus of maximum heating coincide in time and space with
the superoscillation.

Up to this point we have left $Q_{\mu_0}$ in units of the compressive viscosity $\mu_0$. This was done on purpose, to separate the question of whether superoscillations are associated with increased heating from the issue of whether this increased heating is enough to overcome radiative losses and actually increase the coronal plasma temperature. In order to approach this second question it is  important to stress that the heating mechanism associated with superoscillations is very local in nature. Only a small volume of the region of wave propagation, of the magnetic flux tube, is actually heating. And only for a certain time. At any given height along the magnetic field, heating happens during the part of the wave period during which the kink mode substitutes the sausage.  Superoscillations are not heating the whole corona; they are not even heating the whole magnetic flux tube along which the wave propagates. Superoscillations heat small volumes along particular trajectories inside the magnetic flux tube.

Radiative losses were first computed by \cite{Raymond1976}. Subsequent computations improved the precision of the curves but did not add much to the general picture or to the approximate rates. It is curious to notice that radiative losses reach a maximum around $5\times 10^5 K$ and decrease for larger temperatures. This means that any coronal heating mechanism  at constant densities  has  only to  heat plasma up to 1 million K. Any further increase in temperature will reduce radiative losses making the heating mechanism more and more efficient. {This temperature runoff will only stop at around 3million K,when radiative losses do increase again for a short interval of temperatures.} This behaviour of the radiative losses argues against an ubiquitous heating mechanism for the solar corona. Otherwise the equilibrium temperature of the corona would be above 30 million K, and not just around the 1 million mark.

At  this typical equilibrium temperature of $10^6 K$  and assuming densities (both electron and proton) of $10^{9.5} particles/cm^3$ as \cite{Porter1994a,Porter1994} do, \cite{Raymond1976} estimate radiative losses at $10^{-2.81} erg/cm^3 s$.  \cite{Braginskii1965}, on the other hand,  provides expressions for the compressive viscosity  in terms of the ion densities, collision times and  temperature. At $10^{9.5} particles/cm^3$ and   $10^6 K$, the compressive viscosity $\mu_0$ ends up being of the order of $0.1 g/cm s$. In units of $\mu_0$, radiative losses amount to $0.015\times \mu_0$, a number that we can now compare to the heating rate values illustrated in Figs.\ref{Total_Q},\ref{HistoT} and \ref{HistoG} . The value of the density  corresponds to the one used by \cite{Porter1994a,Porter1994}, but the wave amplitudes have been selected so that velocity amplitudes correspond to the observations of coronal waves by CoMP \citep{Threlfall2013,Tomczyk2007}. It is therefore very satisfactory to retrieve in these conditions and with our analytical model the same results as \cite{Porter1994a,Porter1994}: pure magnetoacoustic wave modes, kink or sausage, dissipate due to compressive viscosity, but the resulting heating is not sufficient to compensate radiative losses. But even more satisfactory is the fact that the superoscillation pulse overcomes radiative losses and succeeds in heating the coronal plasma whenever phases are right and the transition happens in less than 0.7 periods (see Fig. \ref{Total_Q}). At higher densities, radiative losses, that increase as density squared, will little by little leave only the strongest superoscillatory pulses as heating sources, at lower densities on the other hand, even the basal wave heating may be enough. But this is beyond the purpose of this paper: our goal here is just to point out that superoscillations, appearing during the transition between wave modes, give rise to a  pulse of heat, localised in time and space.

\begin{figure}
   \centering
   \includegraphics[width=0.45\textwidth]{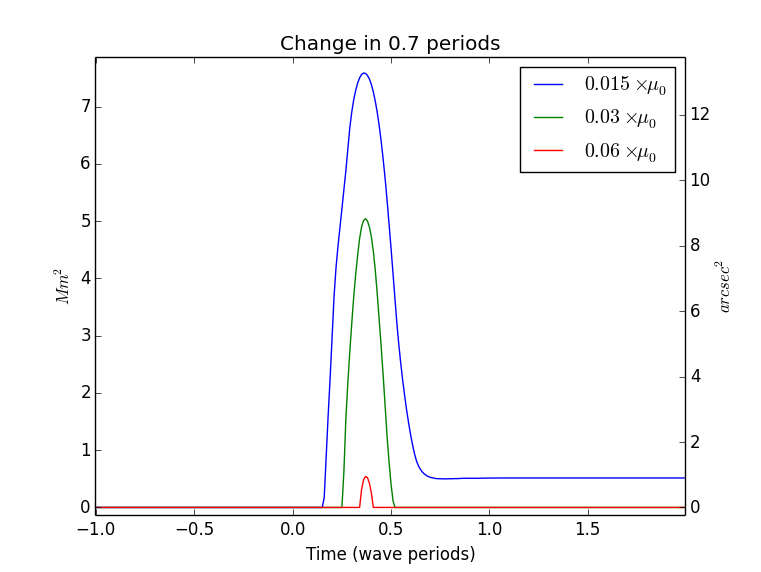}

      \caption{Size of the heating region as a function of time for three different threshold values. The highest heating rates take place in small and localized regions in the neighbourhood of the dislocation. Transverse sizes of 1-2 arcsec (making the area circular for simplicity) are heated. These heated regions propagate along the wave path, both because of phase speed and because of the increased thermal conductivity along field lines. In time, a thin and long region is heated, ressembling 
      bright coronal loops inside the large magnetic flux tube.}
         \label{tam}
   \end{figure}

How big is this heating region inside the magnetic flux tube? Fig. \ref{tam} shows the result of simply adding up those pixels in our discretized volume for which the heating rate exceeds a threshold value. The parameter $m_0$ computed from the imposed sound and Alfv\'en velocities, allows us to convert those pixels into area measurements. Three threshold values are plotted in units of the compressive viscosity $\mu_0$: 0.015, 0.03 and 0.06. The first of those must be at the limit of compensating radiative losses, and, in our particular example, the kink mode actually has regions where heating is larger than this threshold. The heating pulse is here evident in both its short time span (0.7periods at most) and spatial size. The 0.015 threshold is attained at the peak of the heating pulse over a region of $7 Mm^2$. If circular in shape, this would correspond to a radius of $1.5Mm$ or about $2.1\, arcsec$.   This is the maximum extent of the heating region above this threshold. Larger heating rates are attained in smaller regions for shorter amounts of time. Above the $0.03\times \mu_0$ threshold, heating is limited to a time span of 0.3 periods and barely 1.8 arcsec in radius (always assuming a circular shape) .

It is enticing that our heating regions appear to select a thin and long region along the field inside a much larger magnetic flux tube. Long because of two reasons: first, the thermal conductivity is so much larger along the field that any increase in temperature will propagate rapidly along field lines; second, because the transition between the sausage and the kink that gives rise to superoscillations and the heating pulse will also propagate at phase speed along the field lines.
We speculate that this is the reason for the shape of coronal loops: The magnetic field tube is as large as the sunspot umbral sizes, and this transverse scale corresponds to the maximum region of oscillation. But inside such region large heating rates will appear only along thin tubes parallel to field lines and roughly near the center of the kink oscillation. Such hot loops will not be different from the neighbouring corona in either density or magnetic field. This volume of increased temperature will correspond just to the size of the superoscillating region.

\section{Discussion on the approximations made}

\subsection{Chromospheric and coronal waves}

The heating pulse associated with superoscillations analysed in the previous section occurs in coronal conditions (density, phase speeds, etc). But many of the settings of our model are fixed by observations of chromospheric waves. Thus, it is in the chromosphere that the drastic transition between a sausage and a kink (and its associated superoscillation) has been observed. Also the fact that only the $m=+1$ polarisation of the kink mode is excited comes from  observed  chromospheric waves. There is no doubt that kink and sausage modes are found in the corona. Wave dislocations observed by CoMP in coronal loops were explained as a mixture of those two wave modes projected onto the line of sight. But, are these the same waves that are observed in the chromosphere?

The issue at stakes here is that of the chromospheric cutoff. The literature is abundant on the description of the chromosphere as a non-transmissive layer for these waves. And it is also abundant on the conditions on which wave transmission can happen
\citep{Khomenko2015}.  More or less inclined fields have been pointed out as a guide for these waves into the corona. But many works prefer to talk about leakage or even non-linear transformations among different kinds of MHD waves (see again the review by \cite{Khomenko2015} for references to these results).

Our work has implicitly assumed that those chromospheric waves continue their propagation along the field through the chromosphere into the corona unscathed, at least in what concerns their topology (wave mode, dislocations).  If the magnetic field really behaves as a waveguide, even if just a fraction of the original wave amplitude is transmitted, our model stands. However there is a question for further study whether dislocations subsist if the wave is regenerated at the chromospheric boundary by some non-linear process.

\subsection{Fast and slow modes}

All throughout this work, no mention has been done on whether the studied were fast or slow magnetoacoustic waves. For several reasons that we list here, we think it may be misleading to identify whether our waves were fast or slow. And indeed the careful reader may realise that sometimes we refer to observed slow waves and sometimes to observed fast waves, with no discrimination. First, dislocations and superoscillations were observed by \cite{LopezAriste2015} in what most certainly are slow waves propagating along the sunspot magnetic field in the photosphere.  However, in order to set the right amplitudes for the waves in the corona, the velocity amplitudes were compared to the measurements of CoMP \citep{Tomczyk2007} which have been attributed, because of the measurement of a propagation speed larger than the speed of sound, to fast waves. So both slow and fast modes have appeared without distinction in this work.

The presence of the dislocation and the superoscillating region is associated to the transition from sausage to kink, or viceversa, { in the way these two modes were defined in Section 2.2, above, and } independently of whether those are fast or slow modes, as long as both waves are able to interfere. So it does not matter that both phenomena were observed in chromospheric slow modes. It is clear that the very same will happen in fast modes as long as a sausage and a kink modes are present. Nevertheless, and as discussed in the previous subsection, we have assumed that, somehow,  both chromospheric and coronal waves are somehow related. In that sense one would expect therefore that it is slow modes that we are modelling in the corona.  In such a case, why would we constraint our wave amplitudes with those data coming from the observation of fast coronal waves by CoMP?
Our justification is that whatever we are modelling, slow or fast modes, they must not have transverse velocity amplitudes larger than what CoMP observed, otherwise CoMP would have seen those larger amplitudes. Our modelled waves, slow or fast, must have amplitudes comparable or smaller than what CoMP observed. In this sense it is reasonable to constraint our modelled waves so that their amplitude is smaller than or equal to what CoMP observed, independently of whether they are fast or slow.

But there is also another issue at play. \cite{LopezAriste2016}  noticed the presence of dislocations in CoMP data and demonstrated that such dislocations could not be reproduced assuming the presence of a single wave mode: fast or slow, kink or sausage. The only way to reproduce the observed dislocations was to assume the simultaneous presence of both one sausage mode { (in the sense of the definition in Section 2.2)} and either a kink mode or an Alfv\'en wave.   In such a scenario where two waves were simultaneously present, what propagation speed was measured with CoMP?   { Was this the phase speed of a fast wave? Or a combination of the speed of the two waves?. We do not have a clear answer to this question and, because of this, we have avoided identifying whether the waves are fast or slow for the present work.}


\subsection{The frequency and velocity difference between  the kink and sausage modes}

In all our calculations since Eq.\ref{super} we have assumed the same frequency for both the kink and sausage modes. As we have shown, peak heat dissipation around the superoscillation lasts a short time, when the  phases of the waves are right. Making both frequencies equal just makes easier to point the right time when this is going to happen and create our plots accordingly.

{Physically, a difference in the velocities of the two waves (for example if one of them was  a slow one, while the other was a fast one) while keeping the frequencies similar would make the parameter $\Delta$ in Eq. (4) large. The second term of that equation, proportional to $\Delta$ only ensures that the phase and wavelength of the wave change from the slow case to the fast case as one wave substitutes the other and this does not contribute to heating. Superoscillations (and the ensued heating) appeared in the third term of that Eq. (4), and in that term the only impact of $\Delta$ is to change the azimuth of the superoscillation. Since changing the azimuth means changing also the relative phase between the sausage and kink waves and the heating efficiency depended on this relative phase, the presence of a non-zero $\Delta$ will change this efficiency at a given height: some times to increase it sometimes to diminish it. }

Finally this difference may also  bring an undesired effect. In the reference frame that sits on the wave, propagating at the phase speed of the wave, the superoscillation is at a fixed position if both waves have the same phase velocity.   As dissipation takes place, the amplitude of the wave in that region will diminish and the heating will stop when the wave is locally damped. So just the energy of a localized region in the wave will be used in heating the corona.  If, on the other hand, one wave is propagating faster than the other, the superoscillation will move through undamped regions of the wave along the loop. This interesting possibility, interesting from the point of view of dissipating as much wave energy as possible, is made impossible by our assumption of equal frequencies and will have to be abandoned in  future work.

\subsection{Propagation transverse to the field lines}
Perhaps the strongest simplification of our work is to neglect any transverse propagation of the wave, transverse to the magnetic field, even if one expects that conditions (density, field strength) may change faster in the transverse direction, increasing scattering and dissipation of any propagation across field lines as compared to traveling along them.

{ Let us remember that our model contemplates a large flux tube (the size of the sunspot umbra) and that the wave is assumed to be excited simultaneously all across the surface of the umbra. }
It is obvious that transverse propagation is an integral part of the problem { of the wave excitation and propagation in such a large area, at least at photospheric and chromospheric heights.}  Not because equations say so, but because observations clearly indicate it in at least two important manners:  First, present observations of chromospheric and photospheric waves in sunspots never show any indication of the wave radial dependence reaching any near to the first zero of the Bessel function (see e.g. the data in \cite{Centeno2006,Centeno2009} ). A ring of zero oscillation followed by a region oscillating in anti-phase would be a remarkable thing to be observed.  Observed sausage waves show an almost constant amplitude in the transverse plane, as if the first zero of the Bessel function was far away from the sunspot umbra. This is roughly supported by the small values of the $m_0$ parameter.  This means that just a small part of the, assumed homogeneous, magnetic field tube is excited. And therefore transverse propagation of the acoustic mode has no particular impediment. Furthermore, observation of chromospheric waves in high cadence images of sunspots \citep{RouppevanderVoort2003,Bogdan2006,Sych2014}  probably show this transverse propagation although sometimes it may be interpreted as just the result of propagation along tilted field lines \citep{Cho2015}. In such observations, intensity is associated with changes in the plasma density (either hydrogen or ionised Ca) at the particular height at which the spectral line is formed: the observed wave is therefore the  density wave associated with the velocity wave we analyse in this work. But except for differences in amplitude and phase, it is the same wave. What such observations of this density wave at a fixed height show is that the originally excited region is truly small (of about 1-2 arcsec) and it propagates horizontally to cover much of the umbra. 

One curious consequence of this transverse propagation follows.   The position of the singularity is at the coordinate $r=0$ in the pure kink wave. The presence of a sausage mode contribution places the singularity away of this position. The larger the amplitude of the sausage mode, the further from $r=0$ the singularity is. Mathematically, one can think that if the kink mode is absent, the singularity sits at infinity.  As the transition from sausage to kink takes place the singularity approaches from infinity   to its final position at $r=0$. In that virtual movement, it traverses regions that the wave has not yet attained in its transverse propagation at sound speed. The wave travels from $r=0$ towards larger and larger values of $r$, while the singularity travels from infinity towards $r=0$. Only when the two meet the heating pulse can take place. A new condition for heating appears: the transverse propagation must take the wave to  the point where the superoscillation is happening.

\subsection{Radiative losses and viscosity}

As said once and again throughout this paper, the purpose of the present work is to study the superoscillations and to reveal the presence of a heating pulse associated with the superoscillatory region.  We have explored under what conditions this heating is larger than the background heating associated with the gradients in either the sausage or the kink waves. This background heating is insufficient at the typical frequencies of coronal waves to compensate radiative losses.   The heating pulse associated with superoscillations is easily 3-4 times larger than that background heating. But is this enough to overcome radiative losses?

In the main text we have, as much as possible, limited our calculations to the factor multiplying the compressive viscosity 
$\mu_0$.  When necessary we have taken a customary value $\mu_0=0.1 g/cm s$  for this viscosity, the one quoted in \cite{Porter1994a,Porter1994} for a coronal density of $10^{9.5} cm^{-3}$ and compared it with the radiative losses at the temperature of $10^6K$ and same density as computed by \cite{Raymond1976} and others.   

While it is clear that a localised heating 3 or 4 times larger than the background heating may be all that is needed to overcome losses, it must also be clear that taking typical but fixed values for the temperature and density of the corona,  as well as for the derived quantities $\mu_0$ and  radiative losses, is not sufficient. The actual computed heating rates associated with superoscillations must, in future work, be included in a computation that takes into account the changes in temperature, density and radiation associated with that heating. Only after that work is done, will it be possible to  answer the question whether superoscillations may effectively be heating coronal loops.

\section{Conclusion} 

We have studied the presence of superoscillations in solar magnetoacoustic waves and its associated heating under coronal propagation conditions. Superoscillations appear naturally around moving dislocations. The presence of a phase singularity in the wave implies that phase changes in $2\pi$ in any closed path around it. Near the singularity, neighbouring oscillating plasma can have very different phases. If the singularity moves, it  forces these  oscillating regions to change phase rapidly. The local frequency increases enormously due to this effect.   A moving dislocation appears in solar magnetoacoustic waves whenever a kink wave mode is substituted for a sausage wave mode.  This was already observed by \cite{LopezAriste2016} in chromospheric waves on a sunspot umbra. We have now made an analytical model  and applied it to the conditions in coronal loops. Our model confirms both the presence of a moving dislocation and a superoscillatory region around it with local frequencies up to 10 times larger than the nominal wave frequency. For the typical 3-minutes waves in the corona, this means that locally the wave may oscillate in just 10 to 20 seconds.

Such high local frequencies are very localised in time and space. The implicit gradients in the wave due to this suggest the potential presence of strong viscous dissipation. We have set a vertical magnetic flux tube with diameter roughly that of a sunspot umbra  along which waves propagate. We have computed in such scenario the gradients of the wave and the heating due to compressive viscosity. A heating pulse associated with the superoscillation appears whenever the transition between both wave modes is short enough (about half a period). The pulse can be 3 to 4 times more intense than the background heating due to the basic wave dissipation alone, studied by \cite{Porter1994a,Porter1994}. It is also very localized in time and space: While the magnetic flux tube modelled has the size of the umbra, the heated region is barely 1-2 arcsec across. In the direction along the magnetic field heat conductivity is large enough to transfer this heat along field lines, but also the heating pulse follows the wave along field lines as it propagates. Altogether this ressembles very much a bright coronal loop inside a large magnetic flux tube.

Basic order of magnitude estimates show that the pulse heats enough to locally overcome radiative losses. However the exact comparison of both heating rates will have to wait a more detailed study of the plasma evolution under these conditions.

In summary, we have shown that superoscillations can heat coronal plasma at least in coronal loops. The only requirement is a rapid transition from one wave mode to another, a transition due just to the excitation conditions in the photosphere, as already observed in chromospheric waves.


%
%
%
\begin{acknowledgements}
This work has benefited from extensive criticism and discussion from many friends that we are happy to acknowledge here:   F. Ligni\`eres, L. Jouve and J. Ballot (IRAP, Toulouse, France)  that contributed through several discussion meetings and in reading the first version. L. Khomenko, M. Luna  and I. Arregui (IAC, Spain), co-authors of previous papers on the subject of dislocations helped once again to put the work in context. K. Dalmasse (HAO-NCAR, USA) corrected many mistakes and advised on further literature to be read. 
\end{acknowledgements}

\bibliographystyle{aa}

\end{document}